\documentclass{aa}

\usepackage{graphicx}
\usepackage{txfonts}
\usepackage{color}

\begin{document}

\title{Bar-like galaxies in IllustrisTNG}

\author{Ewa L. {\L}okas
}

\institute{Nicolaus Copernicus Astronomical Center, Polish Academy of Sciences,
Bartycka 18, 00-716 Warsaw, Poland\\
\email{lokas@camk.edu.pl}}


\abstract{
We study a sample of bar-like galaxies in the Illustris TNG100 simulation, in
which almost the whole stellar component is in the form of a prolate spheroid. The sample is different from
the late-type barred galaxies studied before. In addition to the requirement of a high enough stellar mass and resolution,
the 277 galaxies were selected based on the single condition of a low enough ratio of the intermediate to long axis of the
stellar component. We followed the mass and shape evolution of the galaxies as well as their interactions with other
objects and divided them into three classes based on the origin of the bar and the subsequent history. In galaxies of
class A (comprising 28\% of the sample), the  bar was induced by an interaction with a larger object, most
often a cluster or group central galaxy, and the galaxies were heavily stripped of dark matter and gas. In classes B and C
(27\% and 45\% of the sample, respectively) the bars were induced by a merger or a passing
satellite, or they were formed by disk instability. Class B galaxies were then partially stripped of mass, while those of class C
evolved without strong interactions, thus retaining their dark matter and gas in the outskirts. We illustrate the
properties of the different classes with three representative examples of individual galaxies. In spite of the different
evolutionary histories, the bars are remarkably similar in strength, length, and formation times. The gas fraction in
the baryonic component within two stellar half-mass radii at the time of bar formation is always below 0.4 and usually
very low, which confirms in the cosmological context the validity of this threshold, which has previously been identified in controlled
simulations. Observational counterparts of these objects can be found among early-type fast rotators, S0
galaxies, or red spirals with bars.}


\maketitle

\section{Introduction}

The origin of galaxy morphology to some extent remains an open question. This is especially true for barred galaxies or
for those with an elongated stellar component. In general, late-type spiral galaxies with central bars are classified
as barred \citep{Buta2015}, but early-type objects with similarly elongated components exist as well
\citep{Baillard2011}. At least two different formation mechanism for bars have been identified. They can be created
either by an inherent instability of thin enough disks \citep{Hohl1971, Ostriker1973}, moderated by the
presence of dark matter halos \citep{Athanassoula2003}, or through tidal interactions with other galaxies
\citep{Noguchi1987, Gerin1990, Miwa1998, Berentzen2004, Lang2014, Lokas2014, Lokas2018} or clusters
\citep{Mastropietro2005, Lokas2016}. The latter mechanism tends to produce bar-like rather than barred
galaxies in the sense that almost the whole stellar component is transformed from a disk to a prolate spheroid, and the
outer stars may in addition be tidally stripped.

The mechanism of tidal bar formation has so far mostly been investigated in detail in controlled simulations of tidal
interactions between galaxies and of galaxies orbiting a cluster. It has been demonstrated that the
mechanism is extremely effective in inducing the bar-like shape when the pericenter of the galaxy orbit in a cluster or
an impact parameter of a fly-by encounter with another galaxy is small enough \citep{Lokas2014, Lokas2016,
Lokas2018, Gajda2017}. The tidal force exerted by the companion is then sufficiently strong to temporarily distort the
disk and trigger the permanent change of stellar orbits into more radial shapes, thus leading to the formation of a stable bar.
If the outer part of the disk is additionally stripped and lost, an elongated stellar component remains. Further
evolution and subsequent tidal shocks at the following pericenters may shorten the length of the bar to make it more
spherical. Because in the cluster environment the gas is also lost due to ram-pressure stripping, the remains appear
like a red elliptical galaxy.

This scenario has recently been placed in the cosmological context \citep{Lokas2020b} by using the
IllustrisTNG simulations of galaxy formation \citep{Springel2018, Marinacci2018, Naiman2018, Nelson2018,
Pillepich2018}. These simulations allow following the evolution of dark matter and baryons in a sufficiently large
region and with good enough resolution to study properties of single galaxies. It has been
demonstrated that the simulations are able to reproduce different observed properties of galaxies, in particular, their
morphologies \citep{Nelson2018, Genel2018, Rodriguez2019}.

\begin{figure*}
\centering
\includegraphics[width=17cm]{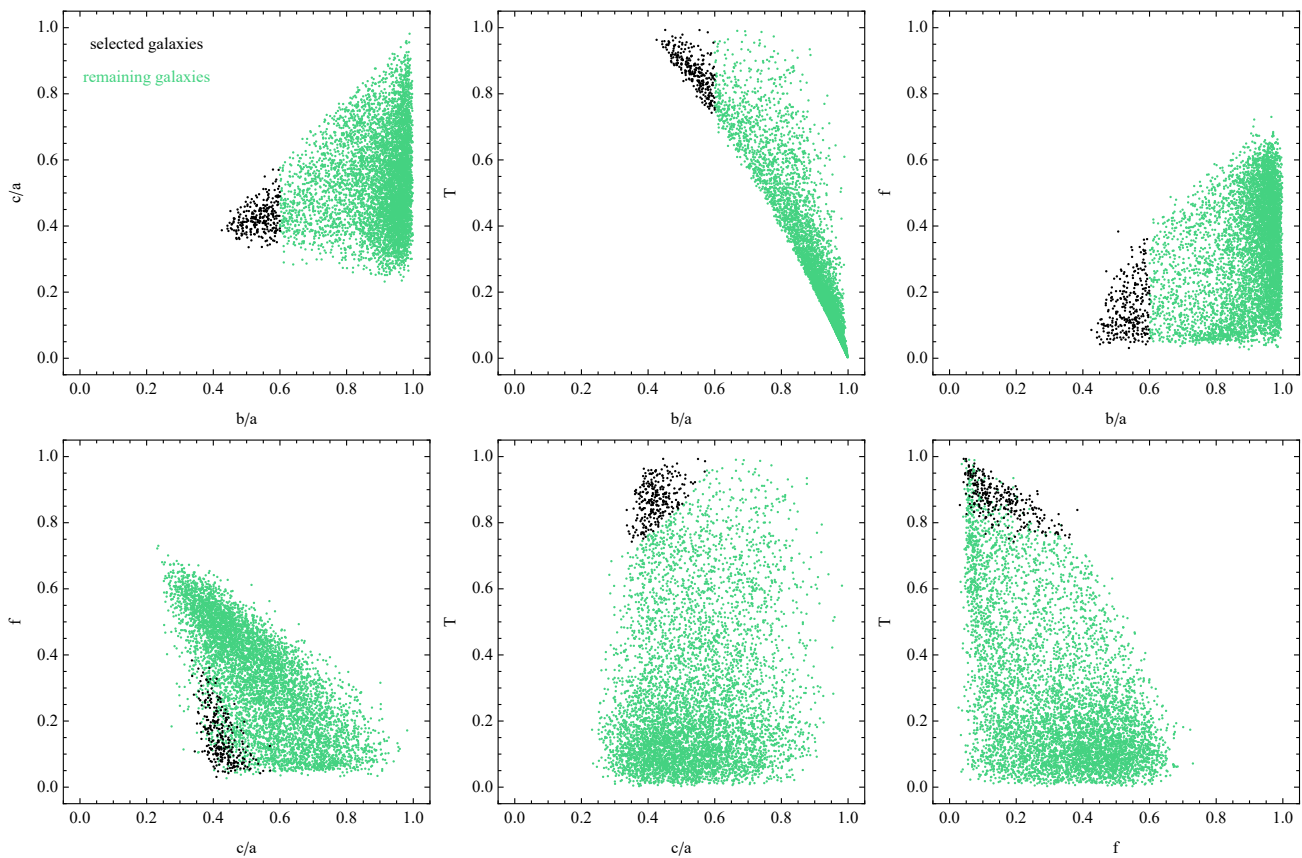}
\caption{Selection of the galaxy sample. The six panels show the properties of the selected
galaxies in different parameter combinations: the axis ratios $b/a$ and $c/a$, the triaxiality parameter $T,$ and
the rotation parameter $f$. Black points correspond to the selected galaxies, and the green points show the remaining
galaxies of the total of 6507 objects.}
\label{selection}
\end{figure*}

\citet{Lokas2020b} demonstrated that the evolution of galaxies in a cluster leads to the formation of tidally induced
bars. A few convincing examples of galaxies were presented in which the formation of a bar coincides with a pericenter
passage on an orbit around the cluster, and in another few galaxies, the bars might have been enhanced by the
interaction with the cluster. In addition, these galaxies were ram-pressure stripped of their gas and had a
substantially reduced dark matter fraction compared to objects that interacted less strongly. However, some galaxies
with very similar properties were also indentified that had never interacted with the cluster. One such case has
recently been described in \citet{Lokas2020c}. Here the bar was induced by a merger and a passing satellite, and the
gas and dark matter were stripped by a subsequent interaction with a group of galaxies. This shows that bar-like
objects can originate in a variety of scenarios.

In this work we study the whole population of bar-like galaxies in the Illustris TNG100 simulation and investigate
their possible formation scenarios. We show that the galaxies have a variety of properties. Some galaxies are dominated
by dark matter and still contain gas; these have not interacted strongly with any neighbors in the past. At the same
time, they are not bars embedded in disks; these have been studied previously using Illustris and IllustrisTNG
simulations \citep{Peschken2019, Rosas2020, Zhou2020, Zhao2020}. In order to distinguish them from the previously
studied objects, we refer to these galaxies as bar-like rather than barred. In section~2 we describe our sample of
selected galaxies. In section~3 we discuss their evolutionary histories, dividing them into three subsamples, and we
present three fiducial examples in more detail. Section~4 describes the properties of the bars in detail, and the
discussion follows in section~5.

\section{Sample selection}

For the purpose of this work, we used the publicly available simulation data from IllustrisTNG described by
\citet{Nelson2019}. In order to have a sufficient number of objects in the sample and enough resolution in each
object, we chose the TNG100 run. We selected the galaxies at $z=0$ by first restricting the sample to subhalos
with total stellar masses greater than $10^{10}$ M$_\odot$, which translates into about $10^4$ stellar particles per
object and thus makes the morphological analysis possible. This requirement is met by 6507 objects in the
final output of the Illustris TNG100 simulation. Next, we imposed a single condition: the intermediate to
longest axis ratio $b/a$ of the stellar component had to be lower than 0.6. For these values we used (and reproduced) the
measurements based on the mass tensor of $b/a$ within two stellar half-mass radii, $2 r_{1/2}$, provided by the Illustris
team in the Supplementary Data Catalogs of stellar circularities, angular momenta, and axis ratios\footnote{The
catalogs are available from the webpage https://www.tng-project.org/data/docs/specifications/\#sec5c.} and calculated
as described in \citet{Genel2015}. The axis ratios were estimated from the eigenvalues of the mass tensor of the
stellar mass obtained by aligning each galaxy with its principal axes and calculating three components ($i$=1,2,3):
$M_i = (\Sigma_j m_j r^2_{j,i}/\Sigma_j m_j)^{1/2}$, where $j$ enumerates over stellar particles, $r_{j,i}$ is the
distance of stellar particle $j$ in the $i$ -axis from the center of the galaxy, and $m_j$ is its mass. The eigenvalues
were sorted so that $M_1 < M_2 < M_3,$ which means that the shortest to longest axis ratio is $c/a = M_1/M_3$, while the
intermediate to longest axis ratio is $b/a = M_2/M_3$.

The sample of galaxies with $b/a<0.6$ contains 288 objects. With this choice, we ensured that  strongly prolate
objects were selected, because by definition the axis ratios obey $c/a < b/a$, where $a$, $b,$ and $c$ are the
longest, intermediate, and shortest axis, respectively. The properties of the selected galaxies in comparison to the
whole sample are shown in Fig.~\ref{selection}. In the six panels of the figure we plot the positions of the galaxies
in different planes of parameters: the axis ratios $b/a$, $c/a$, the triaxiality parameter $T =
[1-(b/a)^2]/[1-(c/a)^2]$, and the rotation parameter, $f$. The rotation parameter is defined as the fractional mass of
all stars with circularity parameter $\epsilon > 0.7$, where $\epsilon=J_z/J(E),$ and $J_z$ is the specific angular
momentum of the star along the angular momentum of the galaxy, while $J(E)$ is the maximum angular momentum of
the stellar particles at positions between 50 before and 50 after the particle in question in a list where the stellar
particles are sorted by their binding energy \citep{Genel2015}.

In the three upper panels of Fig.~\ref{selection} the axis ratio $b/a$ lies in the horizontal axis. The panels illustrate the
selection with the simple cutoff at $b/a < 0.6$. In the lower three panels the positions of the selected galaxies are
non-trivial, and all the selected galaxies have $T > 0.74$ and $f < 0.39,$ which means that they are
indeed prolate and do not rotate fast. We note that $T > 2/3$ is typically considered as characteristic of prolate
objects. However, it is not sufficient to apply this condition for the sample selection because galaxies with high
triaxiality can still have high values of the axis ratios and therefore can be close to spherical. We also note that $f
< 0.2$ is often used \citep{Peschken2019} as a way to select galaxies that are not disky and do not rotate, and most of our
selected galaxies indeed obey this condition.

For all the selected galaxies we determined the evolutionary histories in terms of mass and shape evolution. We
further rejected 11 galaxies whose progenitors could not be traced far enough into the past to
determine the time of bar formation (see section~4), or in which the bar-like shape formed only in the last few simulation
snapshots and therefore could not be considered a stable feature. Our final sample thus contains 277 bar-like galaxies.

\begin{figure}
\centering
\includegraphics[width=7.6cm]{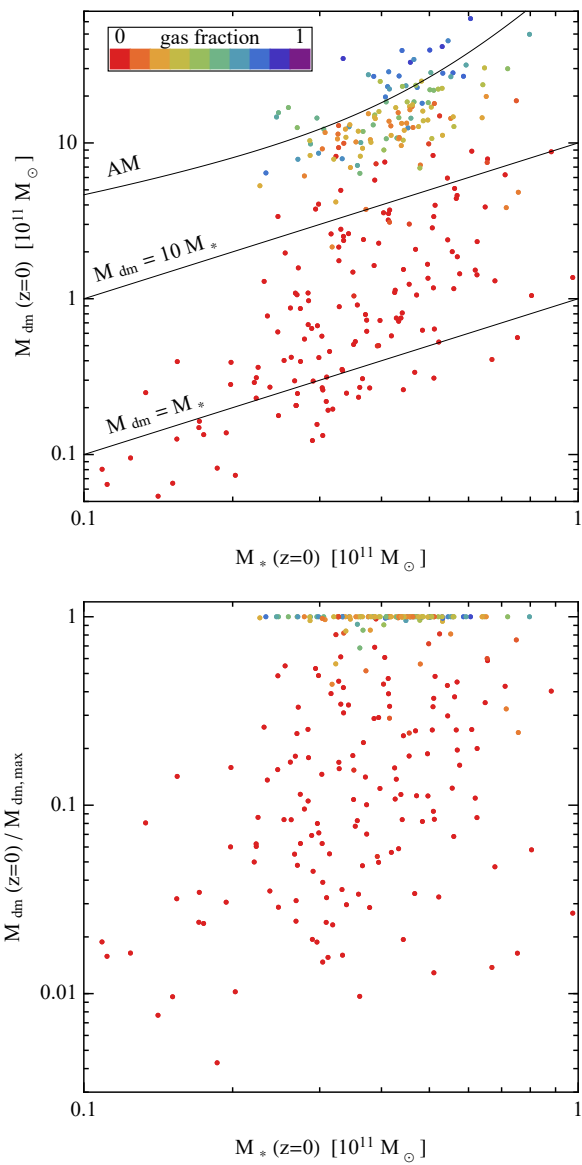}
\caption{Masses of the galaxies in the sample. Upper panel: Dark mass vs. stellar mass of the galaxies at
present. The black lines mark the relations of $M_{\rm dm} = M_*$, $M_{\rm dm} = 10 M_*$ and the prediction
from the abundance matching (AM). Lower panel: Ratio of the present dark mass to the maximum dark mass as a
function of stellar mass. In both panels the points are colored according to the galaxy gas fraction.}
\label{masses}
\end{figure}

Figure~\ref{masses} illustrates the mass distribution of the selected galaxies. In the upper panel we plot the total
dark versus stellar masses of the objects at the present time. The points are color-coded by the gas fraction in the
baryonic component, $f_{\rm g} = M_{\rm g}/(M_{\rm g} + M_*)$. The solid black lines indicate where the dark
masses are equal to stellar, $M_{\rm dm} = M_*$ (lower line) or ten times higher, $M_{\rm dm} = 10 M_*$ (upper line),
as well as a general expectation from the abundance matching for central galaxies \citep{Behroozi2013}. The dark masses are higher than the stellar masses for most galaxies in the sample, although a substantial
number of galaxies (34) contains less dark matter than stars. All the galaxies with $M_{\rm dm} < 10 M_*$ contain
very little or no gas, while galaxies above the threshold, with $M_{\rm dm} > 10 M_*$, typically contain a significant
amount of gas, with the highest gas fraction of 0.87. Overall,
151 out of 277 (55\%) contain gas at present, and the remaining 126 galaxies have no gas at all. We note that the gas distribution is more
extended than the stars, and in particular, the gas fraction within two stellar half-mass radii at present is much
lower, always below 0.12, and higher than 0.01 only for five galaxies.

In the lower panel of Fig.~\ref{masses} we plot the ratio of the present dark matter mass that is bound to the
galaxy to the maximum dark matter mass that the galaxy possessed throughout its history, $M_{\rm dm}(z=0)/M_{\rm dm,max}$,
versus the present stellar mass, $M_* (z=0)$. A significant number of galaxies in the sample lost a large
fraction of their dark mass, probably as a result of tidal interactions with more massive objects. The lowest
ratio of the present to maximum dark matter mass is about 0.4\%. In addition, the galaxies that lost much dark mass
are also gas poor, which indicates that the gas may have been lost by ram-pressure stripping during the same
interactions. Instead, the galaxies with a significant gas content cluster around $M_{\rm dm}(z=0)/M_{\rm dm,max}=1,$
which means that they reach their maximum dark masses at present and are therefore still accreting mass from their neighborhood.

\begin{table*}
\caption{Median parameter values for different samples of bar-like galaxies.}
\label{values}
\centering
\begin{tabular}{c r r c c c c c c c c}
\hline\hline
Sample & $N$ \ \ \  & $M_{\rm dm}/M_*$ & $r_{1/2}$ [kpc] & $g-r$ [mag] & $f_{\rm max}$ & $A_{\rm 2,max}$
& $R_{\rm max}$ [kpc] & $R_{\rm max}/r_{1/2}$ & $t_{\rm f}$ [Gyr] & $f_{\rm g}(<2r_{1/2})$ at $t_{\rm f}$ \\
\hline
Total & 277 \ \ & 10.9 \ \ \ \  & 2.60 & 0.79 & 0.60 & 0.61  & 2.37 & 1.11 & 6.82 & 0.030 \\
A     &  77 \ \ &  1.4 \ \ \ \  & 2.53 & 0.81 & 0.56 & 0.58  & 2.15 & 1.16 & 7.13 & 0.016 \\
B     &  74 \ \ &  5.7 \ \ \ \  & 2.60 & 0.79 & 0.60 & 0.61  & 2.39 & 1.14 & 6.21 & 0.076 \\
C     & 126 \ \ & 35.4 \ \ \ \  & 2.65 & 0.78 & 0.61 & 0.62  & 2.48 & 1.05 & 6.91 & 0.027 \\
\hline
\end{tabular}
\end{table*}

\section{Evolutionary histories with three fiducial examples}

The different mass and gas content properties of the galaxies suggest that their evolutionary histories were varied. To
explore them, we traced the evolution of the different mass components and other properties in time for each galaxy.
The properties that provide most information on the formation of the bar-like shape include the evolution of the axis
ratios and the triaxiality parameter as well as the rotation parameter discussed in the previous section. In
addition to these, we also calculated the evolution of the commonly used measure of the strength of the bar
\citep{Athanassoula1986, Athanassoula2002, Athanassoula2013} in the form of the $m=2$ mode of the Fourier decomposition
of the surface density distribution of stellar particles projected along the short axis. It is given by $A_2 (R) = |
\Sigma_j m_j \exp(2 i \theta_j) |/\Sigma_j m_j$, where $\theta_j$ is the azimuthal angle of the $j$th star, $m_j$ is
its mass, and the sum goes up to the number of particles in a given radial bin. The measurements were first made for
stars within two stellar half-mass radii, $2 r_{1/2}$. The formation of the bar-like shape should be visible as a
decrease in $b/a$, increase in $T,$ and decrease in rotation parameter $f$ (due to the transition from circular to more
radial orbits in the bar). The most direct signature of the bar formation is the increase in bar mode $A_2$, however.

\begin{figure*}[ht!]
\centering
\includegraphics[width=6cm]{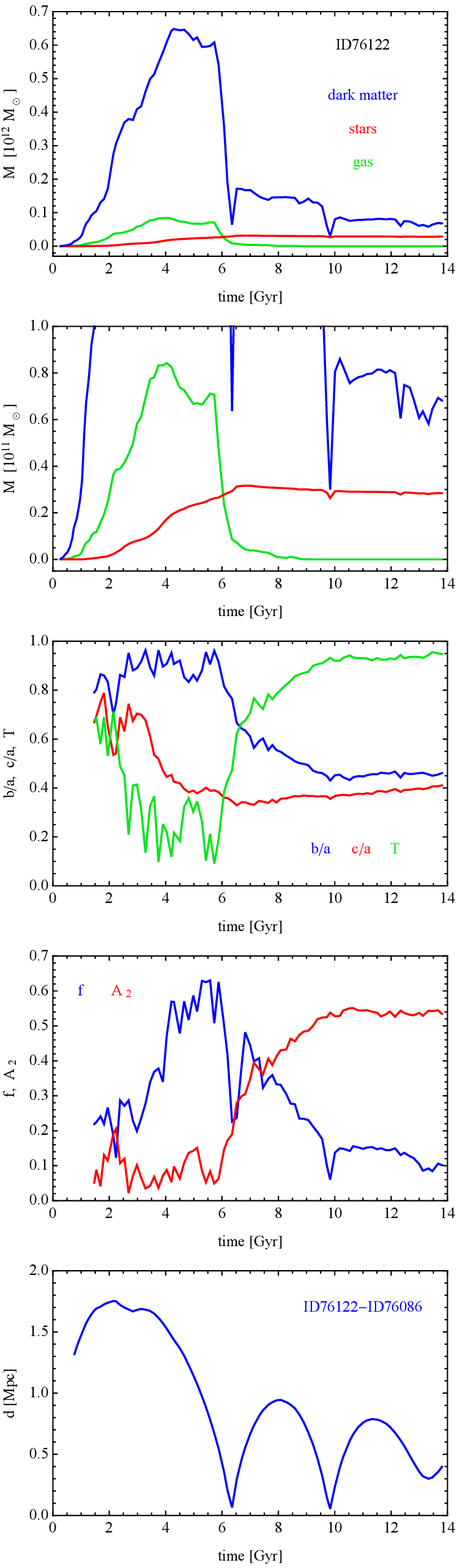}
\includegraphics[width=6cm]{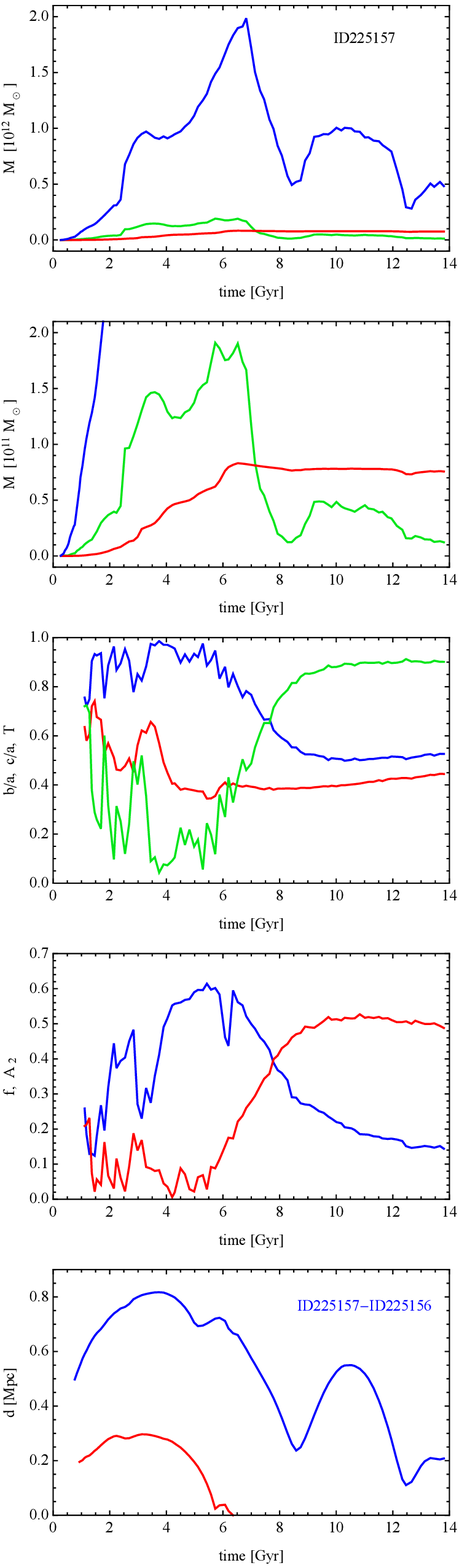}
\includegraphics[width=6cm]{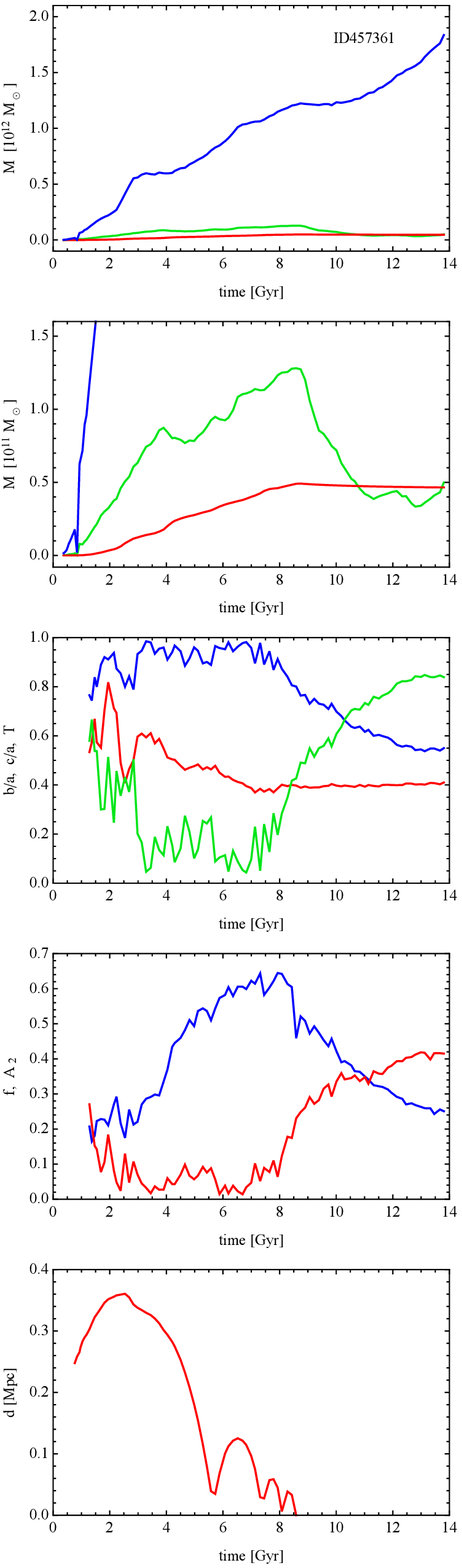}
\caption{Evolution of galaxies belonging to classes A, B, and C. The columns present the results for different galaxies, ID76122
(class A), ID225157 (class B), and ID457361 (class C). Upper row: Evolution of the total dark, stellar, and gas mass
shown with the blue, red, and green lines, respectively. Second row: Same masses, but the vertical scale
is adjusted to the stellar and gas mass. Third row: Evolution of three structural properties of the galaxies, the axis
ratios $b/a$ (blue line) and $c/a$ (red), and the triaxiality parameter $T$ (green). Fourth row: Rotation measure
in terms of the fractional mass of stars on circular orbits $f$ (blue) and the bar strength $A_2$ (red). Fifth row:
Distance of the galaxy from a more massive object (blue) or a merging satellite (red).}
\label{evolution}
\end{figure*}

In addition to tracing the evolution of these parameters, we also verified whether the formation of the bar could be due to
some external factor, such as a merger or tidal interaction with another object. For this purpose we used the
SubLink merger trees of the subhalos \citep{Rodriguez2015} that are provided together with the IllustrisTNG
data, searched for neighbors within 500 kpc of a given galaxy and determined the relative distance between
the galaxy and the neighbor as a function of time. This analysis, and the inspection of evolutionary histories such as
those shown in Fig.~\ref{evolution} below for three fiducial examples, led us to conclude that the sample of bar-like
galaxies can be divided into three subsamples based on their evolution and the origin of the bar-like shape. In the
following we refer to these samples as classes A, B, and C.

We assigned galaxies to class A (comprising 77 objects) if the transition from an oblate to a prolate shape
(i.e., the formation of the bar, see section 4) coincides with a pericenter passage around a more massive
companion, which suggests that their bars were tidally induced by an interaction with a larger galaxy, more often a
central galaxy of a cluster or group (64 objects) than another galaxy (13 objects). Such galaxies lost
most of their dark matter mass and all the gas, and they contain no gas at present (except for two objects with very
low gas fractions).

The galaxies were assigned to class B (74 objects) if their bar had formed differently, by mergers or passing
satellites, or just by a bar instability, but they interacted later with a more massive object and lost a significant
amount of dark mass (more than 50\% of the maximum dark mass). During such interactions, these objects are also often
stripped of their gas so that most of them (51 out of 74) are completely gas-free at present, while the remainder
retain small amounts of gas (with gas fractions up to 0.16).

\begin{figure}
\centering
\includegraphics[width=6.7cm]{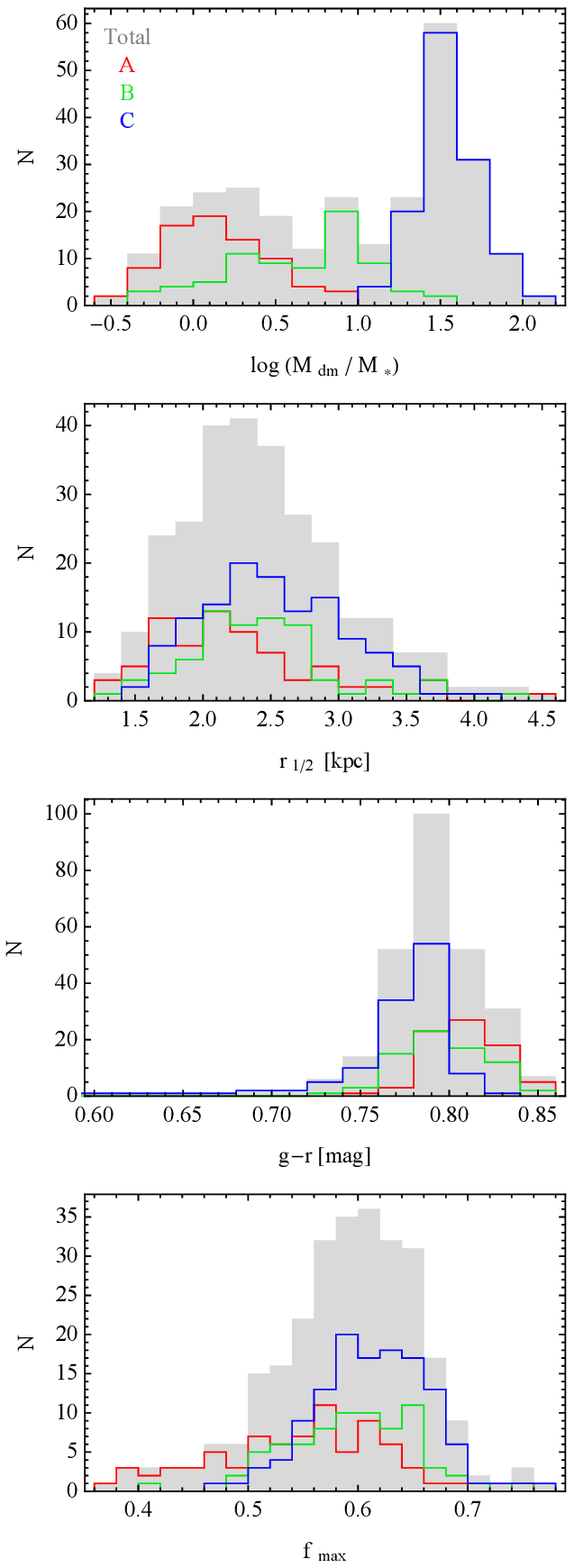}
\caption{Distributions of different properties for the whole sample of bar-like galaxies (gray shaded histograms)
and for classes A, B, and C (colored histograms).}
\label{histograms}
\end{figure}

\begin{figure*}
\centering
\includegraphics[width=5.6cm]{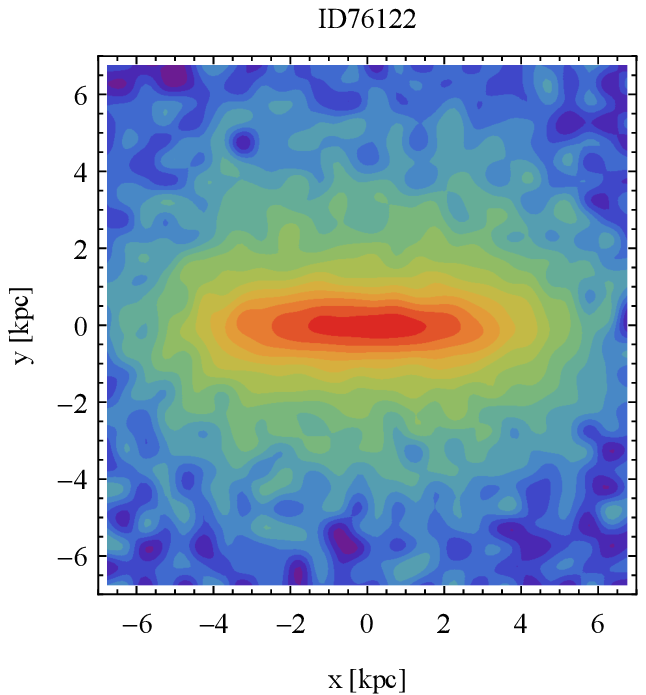}
\includegraphics[width=5.6cm]{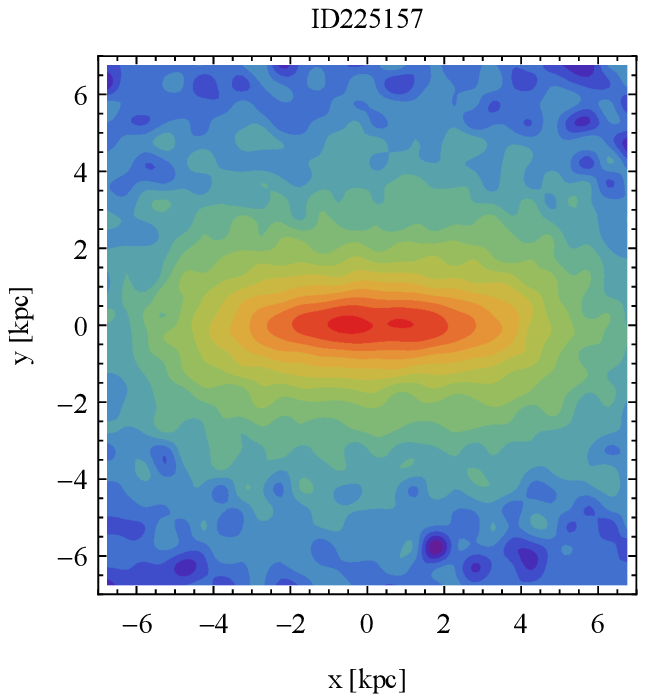}
\includegraphics[width=5.6cm]{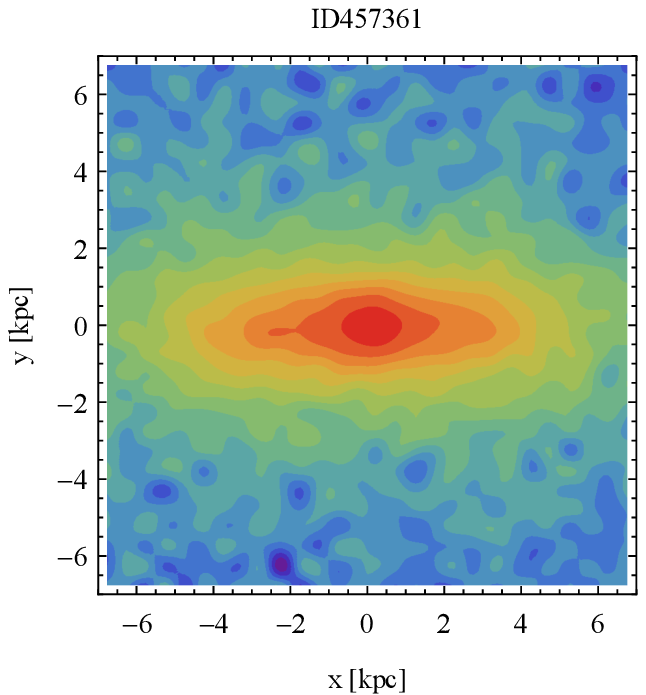}
\caption{Surface density distribution of the stellar components of galaxies ID76122, ID225157, and ID457361
of class A, B, and C, respectively, in the face-on view at the present time. The surface density, $\Sigma,$ is normalized to
the central maximum value in each case, and the contours are equally spaced in $\log \Sigma$.}
\label{surden}
\end{figure*}

The galaxies were included in class C (126 objects) if they did not show any signatures of strong interactions
with more massive objects and did not lose more than 50\% of the maximum dark mass. In contrast to members of classes A
and B, they typically grow in mass and continue to merge with small satellites until the present time. In these objects the
bars form as in class B by mergers or passing satellites or through a bar instability of the disk. All the objects of this class
retain some gas until the end, with gas fractions between 0.01 and 0.87.

Classes A and C can be understood as opposite in terms of the strength of the interactions they experienced and
their effect on morphology. In the case of class A, the interactions were especially strong and resulted in the
morphological transformation from a disk to a prolate spheroid in addition to the very strong mass loss. Class C
galaxies instead evolved rather in isolation, steadily growing in mass or experiencing only very mild tidal effects
from more massive structures, and their morphological transformation was not induced by such effects. The galaxies of
class B are intermediate between A and C in the sense that their tidal interactions were significant in terms of mass
loss, but not directly related to the morphological transformation. This picture is confirmed by verifying the
classification of the galaxies as centrals or satellites given in the Subfind catalogs of the simulation.
While almost all galaxies of classes A and B (except one) are classified as satellites, most objects of class C (110
out of 126, or 87\%) are centrals.

Examples of the evolution of galaxies belonging to samples A, B, and C are shown in the columns of
Fig.~\ref{evolution}. The galaxies were named using their identification numbers in the subhalo catalogs and selected
as the most convincing and clear representative cases of a given class. The first two rows of the panels plot the evolution of
the total masses in three different components (dark, stellar, and gas) in different mass scales. The median values of
the ratio of dark to stellar masses at the end of evolution for different samples are listed in the third column of
Table~\ref{values}. The next two rows of panels in Fig.~\ref{evolution} display the evolution of the shape parameters
$b/a$, $c/a$ and $T$ as well as the rotation parameter $f$ and the bar strength $A_2$. The last row of the panels shows the
distance of the galaxy from the neighbors that most strongly affected its evolution.

Galaxy ID76122 (left column of Fig.~\ref{evolution}), classified as belonging to subsample A, grew in mass until about
6 Gyr, the time of first pericenter passage around the massive galaxy ID76086, a central of a cluster, which caused
strong mass stripping and induced the bar, as indicated by a sudden growth of the triaxiality $T$ and the bar mode
$A_2$, as well as by a decrease in $b/a$ and rotation $f$ around that time. The second tight pericenter at about 10 Gyr
led to further mass loss but left the bar-like shape intact, while the amount of rotation was further decreased. The
gas is lost completely at 9.2 Gyr, when the galaxy approached the second pericenter. The third pericenter is more
distant and has little effect on the galaxy properties. The final dark mass of the galaxy is only 2.4 times higher than
its stellar mass.

The bar in galaxy ID225157 (middle column of Fig.~\ref{evolution}), belonging to class B, was most probably induced by
a merger with a subhalo with ten times lower mass that occurred at 6 Gyr. It later entered an orbit around the larger
galaxy ID225156 (the second largest in the cluster), and the first percenter was at 8.5 Gyr, although it was not very
tight. This and the second, tighter pericenter caused enough stripping to rid the galaxy of most of its dark matter
and gas. Its gas fraction at the end is 0.14 and the dark matter mass is 6.4 times higher than the stellar mass.
Another example of this class of objects, ID44, was discussed in detail in \citet{Lokas2020c}.

The bar in the galaxy ID457361 (right column of Fig.~\ref{evolution}), of class C, was most probably induced by an
orbiting satellite with a ten times lower mass that merged with it at about 8 Gyr. This galaxy did not experience
any significant interactions with larger objects during its lifetime and grew steadily in mass until the present time by
accretion of small subhalos. At the end of the evolution, its dark matter mass is almost 40 times higher than the stellar
mass, and its gas fraction is as high as 0.52.

We summarize the properties of the galaxies in the sample, emphasizing the differences between classes A, B, and
C, in Fig.~\ref{histograms}. The upper panel of the figure shows the distributions of the present dark to stellar masses
$M_{\rm dm}/M_*$, for which the medians are given in the third column of Table~\ref{values}. As expected from the
adopted criteria, the galaxies of class A are not strongly dominated by dark matter, and all have $M_{\rm dm}/M_* < 10$,
while those of class C are strongly dominated, with $M_{\rm dm}/M_* > 10$. The galaxies of class B form a wider, intermediate
distribution. The second panel of the figure shows the distributions of the stellar half-mass radius, $r_{1/2}$.
Although the mass loss in stars is not very strong even in class A, the galaxies of classes A, B, and C tend to have
progressively larger radii, reflecting the fact that galaxies of class A are more tidally truncated. This is confirmed by
the higher median values of this parameter in the fourth column of Table~\ref{values}.

Another important property of the galaxies that was not yet discussed is their color, which we measure as the $g-r$
parameter of a subhalo (in terms of SDSS-like filters). This is shown in the third panel. Because $g-r$=0.6 may be
considered as a value dividing the red from the blue population \citep{Nelson2018}, most of the galaxies in the sample
are red, with $g-r>0.6$, and less so for class C than for class A. (Class C contains only six galaxies and
class B one galaxy with $g-r<0.6$, not shown in the plot.) The median color values are given in the fifth column of
Table~\ref{values}. In line with the colors, we find most of the galaxies to be quenched, with nonzero star formation
rates only for 7 objects in class B and 32 in class C (and values above 3 M$_{\odot}$ yr$^{-1}$ only for one object in
class B and one in class C).

Finally, in order to confirm that the bar-like galaxies of the sample indeed formed from disks, either through
their inherent instability or as a result of interactions, we calculated the maximum values of the rotation parameter
$f$ the galaxy acquired during this history. For this purpose we used the evolutionary histories of this parameter,
such as those shown in the fourth row of panels of Fig.~\ref{evolution} for individual galaxies. Similarly to the
cases shown in the figure, maximum values are always higher than the present values (by a factor of 1.5 to 19) and lie in the
range 0.36-0.78, which means that the galaxies were indeed disks dominated by rotation at some point in their history
before transforming into bar-like shapes with more radial motion. The median maximum values (see
the sixth column of Table~\ref{values}) increase from class A to C, which means that strong tidal
interactions are apparently able to induce bars even in less regularly rotating disks.
We note that the decrease in rotation parameter $f$
is a good indicator of a more general transformation of galaxies from disks to spheroids, resulting in
particular from the evolution in clusters, as discussed in detail by \citet{Joshi2020}.

\section{Properties of the bars}

In this section we study the properties of the bars in the galaxy sample, starting with
the three fiducial examples discussed in the previous section. Figure~\ref{surden} shows the projections of the stellar
density in the face-on view for the three galaxies ID76122, ID225157, and ID457361. The images show that they
indeed have bar-like shapes with pronounced elongations in the central parts. Moreover, the bars have various detailed density distributions, from uniformly elongated (ID76122), to distributions with two density maxima
(ID225157) and with bulge-like centers and elongated outer parts (ID457361) similar to barlenses, that is,
lens-like shapes discussed by \citet{Athanassoula2015} and \citet{Salo2017}.

\begin{figure}
\centering
\includegraphics[width=7.3cm]{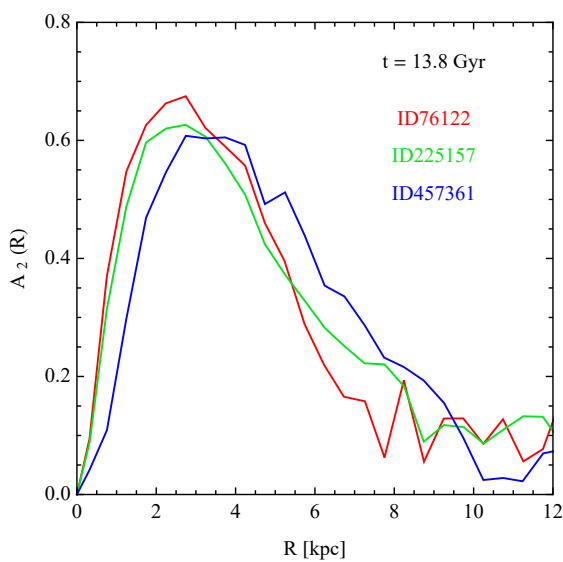}
\caption{Profiles of bar mode, $A_2 (R),$ for galaxies ID76122, ID225157, and ID457361
of class A, B, and C, respectively, at the present time. Measurements were carried out
in bins of $\Delta R = 0.5$ kpc.}
\label{a2profiles}
\end{figure}

The bar-like character of the stellar surface density distribution is further quantitatively confirmed by calculating
the profiles of the bar mode $A_2 (R)$ as a function of the cylindrical radius, $R$. Examples of such profiles for
galaxies ID76122, ID225157, and ID457361 are shown in Fig.~\ref{a2profiles}. Their shapes are typical of bars,
increasing up to a maximum $A_{\rm 2,max}$ at $R_{\rm max}$ and then again decreasing to zero. The value of
$A_{\rm 2,max}$ can be used as a measure of the bar strength, in addition to the global values of $A_2$ within two
stellar half-mass radii plotted in the fourth row of panels in Fig.~\ref{evolution}. The length of the bar can be
estimated as the radius $R$ where $A_2 (R)$ drops to half the maximum value. However, because the $A_2(R)$ profiles are
approximately symmetric around $R_{\rm max}$ and the profiles are typically much more noisy at $R > R_{\rm max}$ than
near the center (because of a small number of stars in the outer bins), it is convenient to use the values
of $2 R_{\rm max}$ as a good approximation of the bar length. For the galaxies ID76122, ID225157, and ID457361, the bar
lengths are thus about 5-6 kpc.

A few galaxies of class A are at present at the pericenters of their tight orbits around more
massive objects and thus interact very strongly. The outer parts of these galaxies are tidally extended, which
manifests itself in a secondary increase in the $A_2 (R)$ profiles at larger $R$. For such objects the values of $A_{\rm
2,max}$ and $R_{\rm max}$ can still be reliably measured, except for one object (ID60783), which has a monotonically
increasing $A_2 (R).$ For this galaxy we performed the measurements using the penultimate simulation output, where
the $A_2 (R)$ is well behaved.

\begin{figure}
\centering
\hspace{0.4cm}
\includegraphics[width=3.5cm]{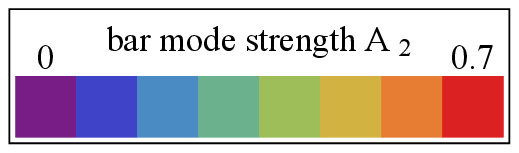}
\vspace{0.1cm}
\includegraphics[width=8.9cm]{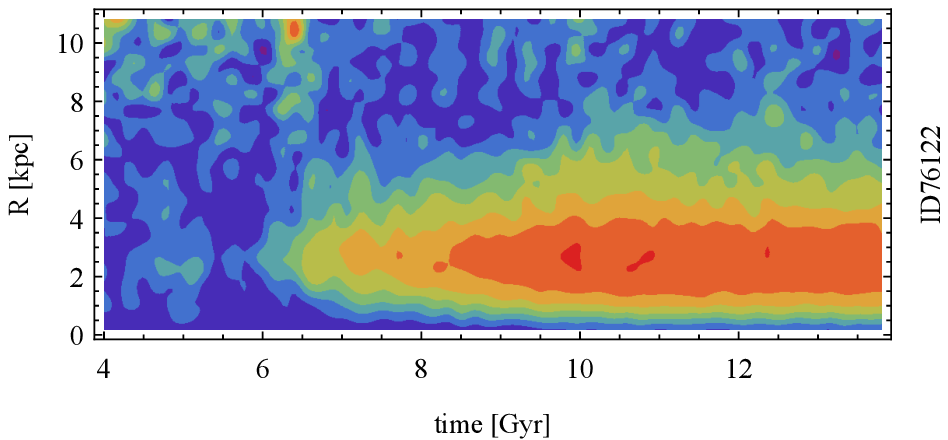}
\vspace{0.1cm}
\includegraphics[width=8.9cm]{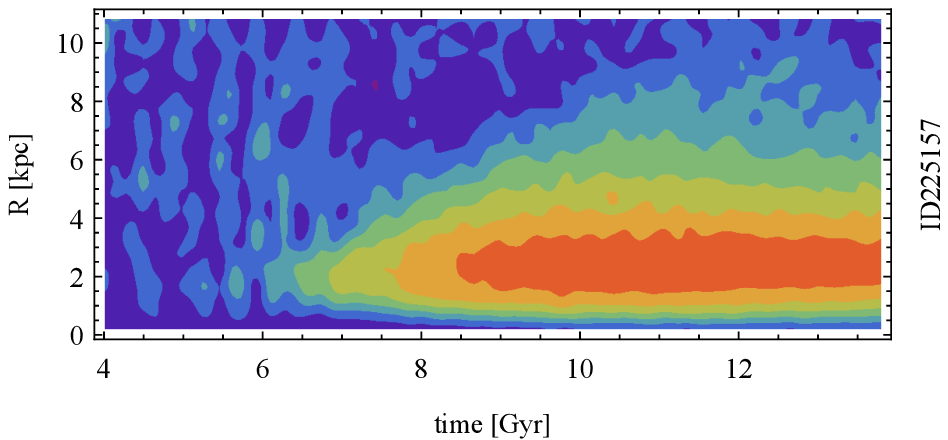}
\vspace{0.1cm}
\includegraphics[width=8.9cm]{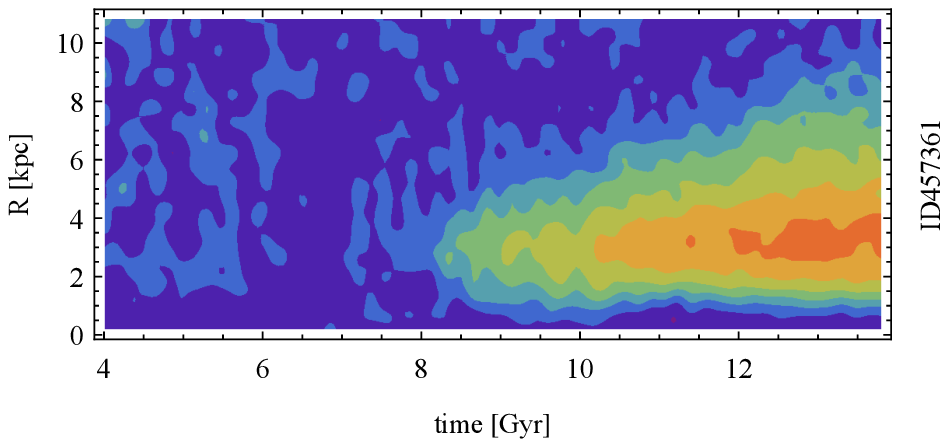}
\caption{Evolution of the profiles of the bar mode, $A_2 (R)$, over time for galaxies ID76122, ID225157, and ID457361
of class A, B, and C, respectively (from top to bottom).}
\label{a2modestime}
\end{figure}

The evolution of the $A_2 (R)$ profiles in time for galaxies ID76122, ID225157, and ID457361 is shown in
Fig.~\ref{a2modestime} in the color-coded form. After they are triggered, the bars grow steadily in strength
and length. For ID76122 (the upper panel of Fig.~\ref{a2modestime}) we note an increase in $A_2 (R)$ at larger $R$
at about 6.3 Gyr, corresponding to the first pericenter passage of this galaxy around a larger host, which induced the
bar. An indication of this increase is also visible at the time of the second pericenter, at 9.8 Gyr. Such effects are easier to see in higher resolution, controlled simulations of the evolution of galaxies in clusters
\citep{Lokas2016}, however.

\begin{figure}
\centering
\includegraphics[width=6.7cm]{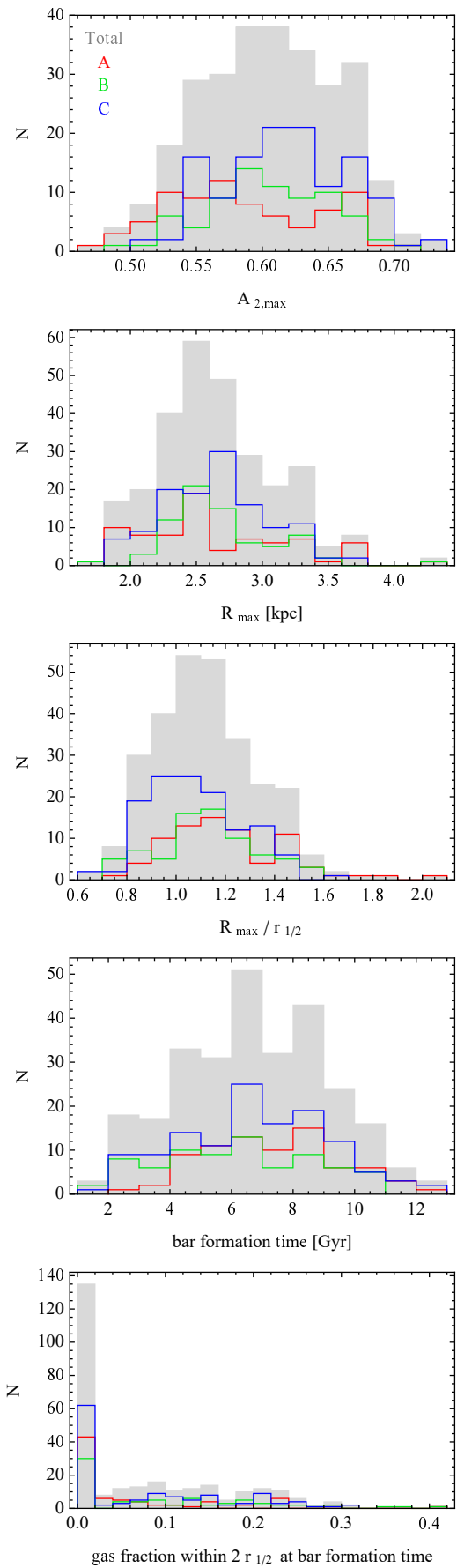}
\caption{Distributions of different properties of the bars for the whole sample of bar-like galaxies (gray shaded
histograms) and for classes A, B, and C (colored histograms).}
\label{histogramsbar}
\end{figure}

We measured the values of $A_{\rm 2,max}$ and $R_{\rm max}$ for all galaxies in the sample. The distributions of these
parameters are shown in the two upper histograms of Fig.~\ref{histogramsbar} for the whole sample and for
classes A, B, and C. Because we  chose strongly elongated galaxies initially using the criterion of axis ratio $b/a
<0.6$, the values of $A_{\rm 2,max}$ are rather high, characteristic of strong bars, but still have a wide
distribution. The values obviously correlate with the alternative measure of bar strength $A_2$, an average bar mode
within two stellar half-mass radii used in Fig.~\ref{evolution}, and another possible measure of ellipticity, $1-b/a$,
although the scatter is large. The seventh and eighth column of Table~\ref{values} list the median values of $A_{\rm
2,max}$ and $R_{\rm max}$ for the whole sample and subsamples A, B, and C, showing that bars
are slightly stronger and longer between subsamples A to C.

The third histogram of Fig.~\ref{histogramsbar} shows the distribution of the ratio $R_{\rm max}/r_{1/2}$, where
$r_{1/2}$ is, as before, the half-mass radius of the stellar component of the galaxy. The values of $r_{1/2}$ are also
systematically higher from A to C (see Fig.~\ref{histograms} and the fourth column of the table) because the
galaxies of classes B and C are not as strongly tidally truncated as those of A, or they are not at all truncated. However, the median values
of $R_{\rm max}/r_{1/2}$  decrease slightly with class, but are always of about unity (the ninth column of
Table~\ref{values}). This means that the length of the bar, estimated as $2 R_{\rm max}$, is typically about
$2 r_{1/2}$.

Next, we estimated the time of the formation of the bar. For this purpose we used the measurements of $A_2$ within two
stellar half-mass radii as a function of time, examples of which are shown in the fourth row of panels of
Fig.~\ref{evolution}. Inspection of these results for many galaxies reveals that although at early times $A_2$ typically
varies strongly, once it crosses the threshold of 0.2, it remains above this value, and the final values for
all galaxies are in the range 0.26 to 0.54. The bar formation times were thus estimated by determining the last
simulation output when $A_2$ was above 0.2 before it dropped below this value when going back in time from the present.
The distribution of the bar formation times estimated in this way is shown in the fourth histogram of
Fig.~\ref{histogramsbar}. The distribution is very wide; some bars formed very early and some only
recently. The typical formation time is of about half the age of the Universe. The median values of this parameter are
listed in the penultimate column of Table~\ref{values} and show that the formation times are earliest for galaxies of
class B. This may be related to the possibility that many of these bars were formed by mergers, which are more frequent
at earlier times.

Finally, we calculated the gas fractions of the galaxies within $2 r_{1/2}$ at the time of bar formation. As stated
above, the lengths of the bars are typically of this size, therefore the gas content in this region is more
important than the overall gas distribution. As we noted before, in the galaxies that contain gas, it tends to be more
extended than the stellar component, but this external gas is expected to have little effect on the formation of the
bar inside $2 r_{1/2}$. The distribution of these inner gas fractions is shown in the last histogram of
Fig.~\ref{histogramsbar}. Interestingly,  most of the galaxies had very little gas in this region at the time of bar
formation. In particular, only six galaxies had a gas fraction between 0.3 and 0.41, and none had a higher value. The
median values of this parameter are listed in the last column of Table~\ref{values}. We note that the value is highest
for the galaxies of class B, which is probably related to the fact that their bar formation times are earlier, so that
the galaxies were naturally more gas rich.

\section{Discussion}

We studied the properties of well-resolved bar-like galaxies in the Illustris TNG100 simulation, selected by the single
criterion of low enough axis ratio of the stellar component, $b/a < 0.6$. After removal of 11 objects that
were not suitable for analysis, the final sample contained 277 galaxies. We traced the evolution of the mass and shape
parameters of the galaxies, which allowed us to divide them into three subsamples A, B, and C based on differences in
their evolutionary histories. The bars in galaxies of class A were induced by an interaction with a larger object, and
the galaxies were strongly stripped of dark matter and gas. The galaxies of classes B and C contain bars that
were induced by a merger or a passing satellite, or that were formed by disk instability. Class B galaxies were then
partially stripped of mass, while those of class C evolved without strong interactions and preserved most of their dark
matter and gas in the outskirts. This classification is in line with the characterization of the galaxies as centrals
and satellites because almost all the galaxies of classes A and B are found to be satellites, while those of class C
are mostly centrals.

Although the bars of galaxies belonging to these samples were created through different mechanisms, their
properties are remarkably similar in terms of strength, length, and formation times. As expected, the bar lengths
correlate well with the stellar half-mass radii of the galaxies, but otherwise, we find no correlations between
different bar properties. For example, there is no correlation between the bar strength and the following parameters:
bar length, stellar mass, formation time, and the gas fraction within $2 r_{1/2}$ at the formation time.

One of the most interesting results of this study is the determination of the upper limit on the gas fraction within $2
r_{1/2}$ at the time of bar formation. In order for the bar formation to be possible, this gas fraction apparently
cannot exceed 0.4. This is consistent with earlier studies using controlled simulations \citep{Shlosman1993,
Athanassoula2013}, although here it was confirmed in the full cosmological context. The exact value of the threshold in
particular agrees very well with our recent study \citep{Lokas2020a}, where we performed controlled
simulations of bar formation in Milky Way-like galaxies with different gas fractions between 0 and 0.4 with a step of
0.1. It was found that only galaxies with gas fractions 0-0.3 formed bars, while those of 0.4 and higher values did
not. These simulations only used a single set of galaxy structural parameters, and the gas distribution followed that
of the stars. However, the results from IllustrisTNG presented here place them in the cosmological context and validate
the threshold of 0.4 as a general threshold because it appears to be obeyed by a variety of galaxies with different
structural parameters that are present in IllustrisTNG.

The bar-like galaxies discussed in this study are different from barred galaxies identified earlier using Illustris and
IllustrisTNG simulations \citep{Peschken2019, Rosas2020, Zhou2020, Zhao2020}. Previous studies focused on bars in late-type fast-rotating disks rather than on the early-type elongated objects we found. The bar-like galaxies
of this study typically contain little gas, and most of the stars contribute to the elongated part rather than a disk.
Although some streaming motion remains in these objects, as is natural for bars, the orbits of stars are elongated
rather than circular, and most of the galaxies have low rotation parameters. The sample of galaxies presented here is
also different from the sample of prolate galaxies in the Illustris simulation discussed by \citet{Li2018}, where the
selected objects were more spherical and bars were explicitly rejected. The sample is also very different from the
prolate rotators discussed by \citet{Ebrova2017}, some of which had prolate shapes.

The bar-like galaxies can have very
different properties in terms of their dark matter content. This reflects their distinct evolutionary paths. The bars
that were formed by tidal interactions with a more massive galaxy (class A) are also heavily stripped of their dark
matter and gas. About half of this class of objects has less dark matter than stars and thus can be classified as
examples of galaxies lacking dark matter recently discussed by \citet{Dokkum2018, Dokkum2019}. The scenario
of tidal stripping, which affects dark matter more than the stars, would be particularly viable for galaxies residing
in groups, as is the case for those observed. The bars in the intermediate class B formed through a different channel
(often by mergers), but the galaxies later interacted with a more massive object and lost some of the mass. The most
numerous class C evolved undisturbed, forming the bar in isolation or in small mergers and retained a high dark matter
content. This class often also contains a significant amount of gas, but mostly in the outer parts, outside the bar.

It is interesting to ask what fraction of the bar-like galaxies discussed here formed by interactions. Unfortunately,
this quantity is very difficult to estimate. Only for class A, which comprises 77 out of 277 galaxies (28\% of
the sample), can we be reasonably certain that the bar was induced by tidal interactions because its formation
coincides with a pericenter passage around a more massive object, and in this configuration, the tidal forces are
particularly strong. For many of the class B and C objects we were able to identify significant mergers coinciding with
the morphological transformation that probably caused enough distortion in the disk to induce a bar. However, the
history of many C class galaxies is very quiescent, and only small subhalos were accreted continuously, and it is
impossible to determine whether any of them contributed to the formation of the bar. Because the formation of the bar is likely a
stochastic process \citep{Sellwood2009}, even a small perturbation may be enough to trigger it.

We may wonder what the counterparts of these bar-like galaxies are among real objects. Some of them, mostly those of
class A, could be indistinguishable from elliptical galaxies. Depending on the resolution of the observations
and the line of sight, they would be classified as fast or slow rotators \citep{Emsellem2007}. The detection of
a significant rotation signal is most probable in the case of the end-on view because the streaming motion of the stars
in the bar is strongest in this case. When viewed edge-on, they may look like disks as a result of their elongation and
rotation. Because they lack disks and spiral arms, they would not in general be seen as typical barred
galaxies. They could be classified as S0s when seen face on, however. Good examples of such objects are PGC 31198 (NGC
3266) and PGC 41302 (NGC 4479) from the EFIGI catalog of nearby galaxies, shown in fig. 15 of \citet{Baillard2011}. NGC
4479 is a member of the Virgo cluster, therefore its bar may have been tidally induced by the interaction with the
central galaxy. Galaxies of class C bear some similarity to red spirals \citep{Guo2020}, many of which possess large
bars and gas only in the outskirts, as well as to the blue compact dwarf NGC 2915 \citep{Meurer1996}, which has a bar-like
stellar component and an extended gaseous disk with almost no stars.

 \citet{Pulsoni2020} recently studied the sample of early-type galaxies (ETG) in IllustrisTNG and identified among them a
subsample of bar-like galaxies with a stellar mass axis ratio $b/a < 0.6$ within one stellar half-mass radius, $r_{1/2}$.
This subsample overlaps our sample to some extent. Our sample was selected by the same condition, but applied to stars
within $2 r_{1/2}$. In particular, we find that out of 277 galaxies in our sample 201 (73\%) also have $b/a < 0.6$
within $r_{1/2}$. After measuring the kinematic properties of their bar-like galaxies, \citet{Pulsoni2020} concluded that
many of them are elongated slow rotators that are not found in samples of real ETGs, nor in the original Illustris
simulation. They suggested that the simulated bar-like galaxies in IllustrisTNG are probably failed disks resulting from the
changed galaxy formation model in TNG with respect to Illustris.

We confirmed that the criterion of $b/a < 0.6$ within $2 r_{1/2}$ indeed
selects only 35 objects in the Illustris-1 simulation, but a comparison similar to the one shown in Fig.~\ref{selection}
reveals that Illustris-1 galaxies are typically thicker and thus less susceptible to bar instability. In general, the
TNG galaxies agree better with observational properties \citep{Nelson2018}, except for overquenching
\citep{Angthopo2020, Sherman2020}, which is most probably due to too strong kinetic feedback from active galactic nuclei, combined with high
thresholds for the seed black hole mass and halo mass. This may be of concern in the case of galaxies of our class C, which
may rid themselves of the gas too efficiently and form bars too easily. It should not be a problem for galaxies
of classes A and B, however, for which the quenching is mostly environmental and the tidal interactions and ram-pressure stripping
of the gas depend very little on the details of the model. These galaxies are clear examples of bars that were tidally
induced or whose growth was aided by the low gas and dark matter content caused by interactions.

While the simulations are certainly not yet perfect, some improvements may also be needed on the observational
side. The classification of the galaxies in terms of their shapes includes deprojection, which is a highly degenerate
procedure for triaxial systems, and in contrast to common belief, the intrinsic shapes of galaxies are not well known. For
example, \citet{Weijmans2014} noted that for slow rotators from the Atlas3D project the prolate shapes provide a very good
fit when there is no intrinsic kinematic misalignment, which is the case of bar-like galaxies. \citet{Bassett2019} warned
that assuming kinematic misalignment, as is typically done for triaxial systems, is not a reliable approach because
there are prolate objects without misalignment. It is therefore possible that more bar-like galaxies will be identified
in observations when the methods of inferring the intrinsic shapes of early-type galaxies improve.

\begin{acknowledgements}
I am grateful to the anonymous referee for useful comments, to Gerhardt Meurer for letting me know about NGC
2915 and to the IllustrisTNG team for making their simulations publicly available.
\end{acknowledgements}

\end{document}